\pdfoutput=1
\documentclass[aps,prb,twocolumn,floatfix,longbibliography]{revtex4-2}
\usepackage{graphicx}
\usepackage{amsmath,amssymb,amsthm}
\usepackage{bm}
\usepackage{xcolor}
\usepackage[colorlinks=true,linkcolor=blue!60!black,citecolor=blue!60!black,urlcolor=blue!60!black,pdftitle={Boundary-controlled algebraic vorticity order in Shakti artificial spin ice},pdfauthor={Chaeyeon Kim, Suhwan Yoo, Jeehoon Choi, Hee Young Kwon, Changyeon Won}]{hyperref}

\newtheorem{theorem}{Theorem}
\newtheorem{proposition}{Proposition}
\newtheorem{lemma}{Lemma}
\newtheorem{conjecture}{Conjecture}
\newcommand{\eps}{\varepsilon}

\newcommand{\floor}[1]{\lfloor #1\rfloor}

\begin{document}

\title{Boundary-controlled algebraic vorticity order in Shakti artificial spin ice}

\author{Chaeyeon Kim}
\author{Suhwan Yoo}
\author{Jeehoon Choi}
\author{Hee Young Kwon}
\author{Changyeon Won}
\email{cywon@khu.ac.kr}
\affiliation{Department of Physics, Kyung Hee University, Seoul 02447, Republic of Korea}

\date{\today}

\begin{abstract}
In the ground states of Shakti artificial spin ice the unhappy vertices form a dimer covering, and each of them joins two perpendicular plaquettes into an L-shaped tetromino. We show that the circulation of the island moments around each tetromino, a vorticity $v=\pm1$, is a hidden collective variable fixed by the dimer height $h$: $v=s(-1)^{\floor{h/4}}$ with a period-4 prefactor $s$. Consequently the island coupling rule is never frustrated, $v$ is unique up to a global sign, and the total vorticity is a topological invariant set by the sample boundary. Averaged over the ground states, $v$ forms a $2\times2$ superstructure with algebraic order whose amplitude decays as the inverse fourth root of the sample size; exact Kasteleyn calculations confirm this and suggest a closed product formula for the amplitude at the center of an Aztec-diamond sample. All vorticity patterns are projections of a single phase field set by the mean height, so the shape, size, and edge termination of a sample select which plaquettes order and with which relative signs. Straight-edged arrays behave as Aztec diamonds below a crossover temperature that decreases with their size. Half vortices are confined by strings, whereas full vortices are free, and dipolar interactions preserve the algebraic order down to a locking transition. The results provide rules for designing and reading out vorticity order in nanomagnet arrays.
\end{abstract}

\maketitle

\section{Introduction}

Artificial spin ices are arrays of elongated single-domain nanomagnets whose moments behave as Ising spins pinned to the edges of a lattice~\cite{Wang2006,Nisoli2013RMP,Skjaervo2020,Sultana2025}. Because the lattice geometry can be chosen at will and every moment can be imaged, they have become a laboratory for frustration, emergent gauge fields, and magnetic charges~\cite{Moller2006,Castelnovo2008,Henley2010,Perrin2016,Farhan2013}. In the square and kagome geometries frustration lives at the vertices, and the low-energy physics is controlled by an ice rule. A second route, \emph{vertex frustration}, was opened by lattices in which not all vertices can simultaneously reach their lowest-energy configuration~\cite{Morrison2013,Nisoli2022,Crater2024}. The best studied example is the Shakti lattice~\cite{Chern2013,Gilbert2014,Lao2018}. Its minimal loops are rectangular plaquettes, and in every ground state each plaquette contains exactly one ``unhappy'' three-leg vertex that it shares with a neighboring, perpendicular plaquette. The unhappy vertices therefore form a dimer covering of the square lattice of plaquettes, which explains the extensive degeneracy, the emergent ice rule, and the classical topological order observed in Shakti arrays~\cite{Chern2013,Gilbert2014,Lao2018}.

The dimer description tells us \emph{where} the frustration sits, but it does not say which collective magnetic pattern the arrays actually display. In the present work we identify such a pattern. We regard each dimer together with its two plaquettes as an L-shaped tetromino (Fig.~\ref{fig:mapping}). The ten islands on the perimeter of a tetromino nearly close a flux loop: nine point along a common sense of circulation and one, the \emph{defect island}, points against it. The tetromino thus carries a vorticity $v=\pm1$ (counterclockwise or clockwise), a natural observable in nanomagnet arrays where flux closure is energetically favored and directly imaged. We ask how these vorticities are organized globally, how the sample boundary controls them, and what happens under realistic interactions and defects.

Our main findings are as follows. \emph{Vorticity is slaved to the height field.} The tetromino vorticity is an explicit local function of the Thurston height of the dimer covering (Sec.~\ref{sec:height}). As a consequence the island coupling rule between tetrominoes is never frustrated, the vorticity of every tetromino is fixed once the tiling is fixed, and the total vorticity $V=\sum v$ is conserved by all local rearrangements. $V$ is a topological invariant of the sample, and we give it in closed form in terms of the boundary alone.

\emph{Algebraic, boundary-imprinted order.} Averaged over the degenerate ground states, $v$ forms a $2\times2$ superstructure whose amplitude decays as $(\text{size})^{-1/4}$ (Sec.~\ref{sec:order}). The exponent follows from the Gaussian free field description of dimers~\cite{Kenyon2001}; we confirm it with exact Kasteleyn calculations and conjecture a closed product formula for the central amplitude of the Aztec-diamond sample.

\emph{One phase field controls all patterns.} All observed $\langle v\rangle$ patterns are projections of a single complex field $\Psi$ whose phase is $\pi\langle h\rangle/4$ (Sec.~\ref{sec:phase}). Sample size modulo 4 and the way the sample is cut from the lattice therefore decide whether the horizontal plaquettes, the vertical plaquettes, or both carry the order; at the sample center the phase is fixed by symmetry up to a sign. Holes and pinned regions act as additional boundaries that reinforce or twist the pattern. Straight-edged arrays are Aztec diamonds, and in arrays cut along block boundaries the imprint survives only below a crossover temperature that decreases with size (Sec.~\ref{sec:axis}).

\emph{Experimental fingerprints and robustness.} The order gives power-law superlattice peaks in the magnetic structure factor (Sec.~\ref{sec:corr}). Section~\ref{sec:robust} summarizes how it responds to defects and to the dipolar interaction and how it can be read out from images: half vortices are confined by strings, full vortices are free and screen the order, and dipolar interactions leave the algebraic phase intact down to a temperature below the vertex gap, where they lock it into long-range order. Section~\ref{sec:conclusion} concludes with design rules. Proofs and methods are collected in the Appendices.

\section{From Shakti ground states to domino tilings}
\label{sec:mapping}

\subsection{Tetrominoes and dominoes}

We use the Shakti lattice in the form of a basketweave of $1\times2$ rectangular plaquettes: $2\times2$ blocks made of two parallel plaquettes, with the block orientation alternating like a checkerboard. Islands sit on the unit edges, except on the middle edge of each plaquette, which is absent. Vertices have coordination $z=2$, 3, or 4.

\begin{figure*}[t]
\centering
\includegraphics[width=\textwidth]{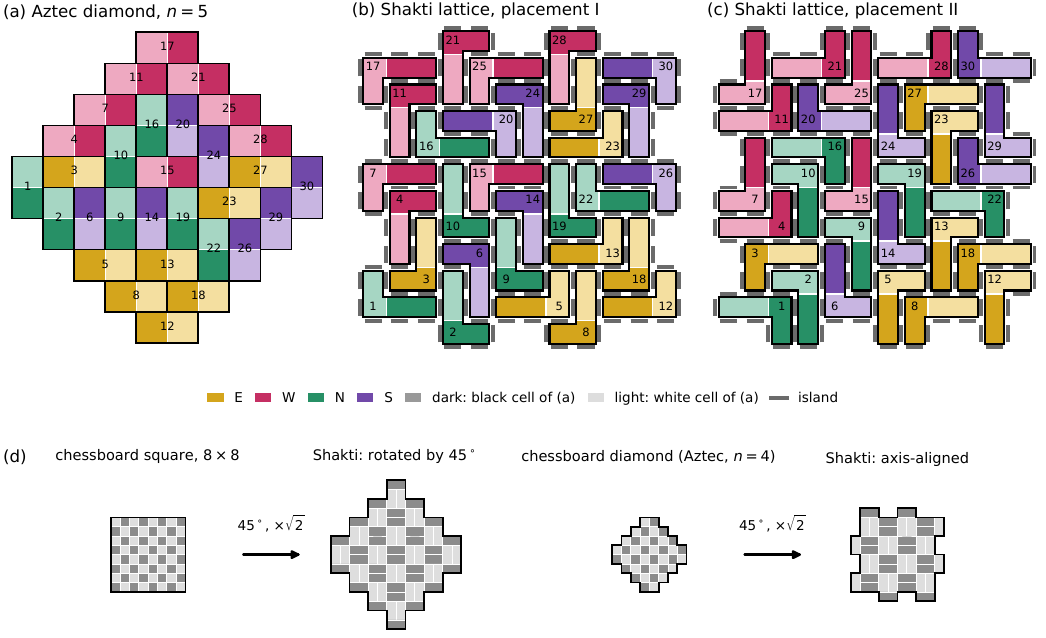}
\caption{Shakti ground states are domino tilings: the two plaquettes that share an unhappy vertex form an L-tetromino, which maps to a domino of the chessboard. (a) A domino tiling of the Aztec diamond of order $n=5$. Colors label the four domino types (E, W, N, S: position of the white cell relative to the black cell); dark and light shades mark black and white cells. (b) The corresponding L-tetromino tiling of the Shakti lattice in placement I. Each tetromino is outlined in black and slightly shrunk, so that the gray bars (nanoislands) lie in the gaps between neighboring tiles; a thin white line separates its two plaquettes. Each domino becomes an L made of two perpendicular plaquettes that share the unhappy vertex. (c) The same tiling in placement II, where the diamond is shifted by one cell relative to the plaquette blocks. In (a)--(c) hue and shade are those of (a): the hue is the domino type, dark (light) marks the black (white) cell of (a), and the number identifies the same domino. In placement I the black cells of (a) become horizontal plaquettes and the white cells vertical ones. The one-cell shift of placement II inverts this, so in (c) the dark part of every tetromino is its vertical plaquette; equivalently, relative to the chessboard coloring of placement II the domino types are those of (a) with E$\leftrightarrow$W and N$\leftrightarrow$S. (d) Geometry of the map, drawn to scale: a chessboard square becomes a square rotated by $45^\circ$ in the Shakti lattice, and a chessboard diamond (here the Aztec diamond of order 4) becomes a square with edges along the Shakti axes. Dark (light) cells and plaquettes are black (white), that is, horizontal (vertical).}
\label{fig:mapping}
\end{figure*}

In a ground state every plaquette contains one unhappy $z=3$ vertex, located at the corner where it touches a perpendicular plaquette~\cite{Morrison2013,Chern2013,Lao2018}. Pairing each plaquette with the plaquette that shares its unhappy vertex gives an L-shaped tetromino made of four unit squares. We map every plaquette to a chessboard cell: horizontal plaquettes to black cells and vertical plaquettes to white cells, such that orthogonally adjacent cells correspond to perpendicular plaquettes that share an edge. The chessboard translations $(2,2)$ and $(2,-2)$ act on the Shakti lattice as translations by $(0,4)$ and $(4,0)$, so the map is a $45^\circ$ rotation with a factor $\sqrt2$, fixed by a fundamental domain of eight cells (Appendix~\ref{app:proofs}). Each cell $c$ also has a diagonal partner $p(c)$, the parallel plaquette in the same block. Under this map a tetromino is a domino and a Shakti ground state is a domino tiling of the corresponding region [Figs.~\ref{fig:mapping}(a) and \ref{fig:mapping}(b)]. The dimer map of Lao \emph{et al.}~\cite{Lao2018} is the same correspondence written for the unhappy vertices; here it is lifted to the level of tiles.

The two plaquettes of a tetromino are inequivalent. The \emph{strong} (short) arm contains the elbow square of the L; the other is the \emph{weak} (long) arm. In chessboard terms the strong cell of the domino $\{x,y\}$ is $y$ if $y$ is orthogonally adjacent to $p(x)$ and $x$ otherwise.

\subsection{Tetromino template, vorticity, and the island rule}

\begin{figure*}[t]
\centering
\includegraphics[width=\textwidth]{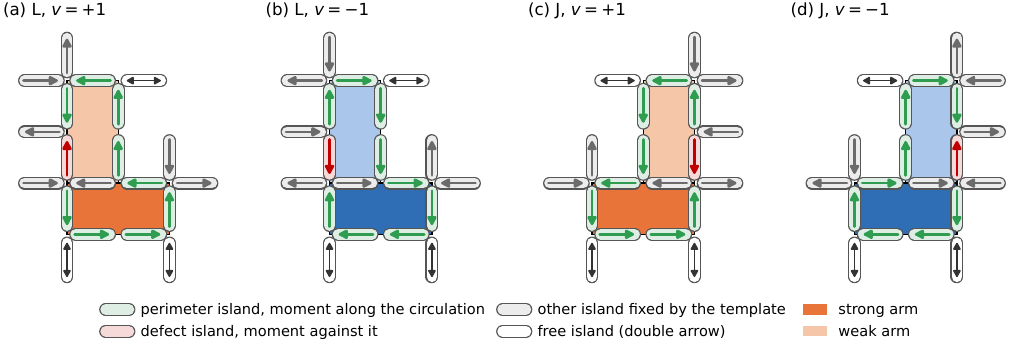}
\caption{The ground-state template fixes all ten perimeter islands of a tetromino, so each tetromino carries a vorticity $v=\pm1$. Templates are shown for both chiralities (L and J) and both vorticities. Each island is drawn as a capsule with its moment as an arrow. Green capsules are perimeter islands whose moments follow the circulation, and the red capsule is the defect island, whose moment points against it. Grey capsules are the other islands the template fixes (the one between the two arms and those just outside the tile); white capsules with double arrows are islands left free. Dark (light) shading marks the strong (weak) arm.}
\label{fig:template}
\end{figure*}

Figure~\ref{fig:template} shows the island configuration of a single tetromino in a ground state. All ten perimeter islands are fixed. Nine of them point along a common circulation, which defines the vorticity $v=+1$ (counterclockwise) or $v=-1$ (clockwise); the tenth, the defect island, points against it. Its position is fixed by the shape: it lies on the weak arm, on the long side facing away from the block partner, on the unit segment next to the elbow. With this template all $z=4$ vertices are of type I and all $z=3$ vertices carry charge $\pm1$. The clockwise state is the time reverse, and the mirror-image (J) tetromino is obtained by reflection. The template also fixes several islands just outside the tetromino.

Two tetrominoes that share an island must agree on its orientation. Since counterclockwise circulations of two adjacent tiles traverse their common edge in opposite directions, this gives a simple rule.

\begin{lemma}[Island rule]
Two tetrominoes that share an island have opposite vorticity, unless the shared island is the defect island of one of them, in which case they have the same vorticity. Two defect islands never coincide.
\end{lemma}

In particular, strong arms never carry a defect island, so tetrominoes in contact through their strong arms always have opposite vorticities, as do block partners (parallel weak arms). The externally fixed islands add further couplings between tiles that touch only at a corner; all of them favor equal vorticities and are consistent with the rule (all coupled placements are checked in Lemma~\ref{lem:local}).

\subsection{Sample geometries}

Artificial spin ices are finite samples, and their boundaries turn out to matter. We consider three families of samples: (a) the Aztec diamond of order $n$, the cell set $|i+\tfrac12|+|j+\tfrac12|\le n$, which is the classic setting for dimers with a frozen boundary and an arctic circle~\cite{EKLP1992,Jockusch1998,Cohn1996}; (b) $L\times L$ square and $W\times H$ rectangular boards; (c) arbitrary simply connected shapes, with or without holes. An Aztec diamond can be cut from the Shakti lattice in two inequivalent ways (placements I and II), which differ by a one-cell shift of the chessboard relative to the plaquette blocks [Figs.~\ref{fig:mapping}(b) and \ref{fig:mapping}(c)]; for square boards the analogue is the color of the corner cell. Since any shape can be fabricated lithographically, we treat all placements on an equal footing.

The map rotates the chessboard by $45^\circ$ and stretches lengths by $\sqrt2$ [Fig.~\ref{fig:mapping}(d)]. A chessboard square or rectangle therefore appears in the Shakti lattice as a square or rectangle rotated by $45^\circ$, with staircase edges, whereas a chessboard diamond such as the Aztec diamond appears as a square with edges along the Shakti axes. An $L\times L$ chessboard square has Shakti side length $\sqrt2L$, and an Aztec diamond of order $n$ fills a $(2n+2)\times(2n+2)$ Shakti square (Sec.~\ref{sec:axis}). Throughout, ``square'' and ``rectangular board'' refer to the chessboard. Shakti arrays with straight edges along the lattice axes, the usual fabricated geometry, are therefore Aztec diamonds (Sec.~\ref{sec:axis}).

\section{Vorticity as a function of the height field}
\label{sec:height}

\subsection{Height formula}

The Thurston height $h$ of a domino tiling lives on the lattice vertices~\cite{Thurston1990}. Along an edge that is not crossed by a domino, $h$ increases by $1$ if the cell to the left is black and decreases by $1$ otherwise; across a domino it changes by $\mp3$. We fix the additive constant so that $h(x,y)\equiv r(x,y)\pmod 4$, with $\bar x=x\bmod 2$, $\bar y=y\bmod2$, and
\begin{equation}
r(x,y)=\bigl[2+\bar x-\bar y+2\bar x\bar y\bigr]\bmod 4 .
\label{eq:residue}
\end{equation}

\begin{proposition}[Height formula]
Let $d$ be a domino whose lower-left cell has lower-left corner $p_d=(x,y)$. Up to one global sign,
\begin{equation}
v(d)=s(d)\,(-1)^{\floor{h(p_d)/4}},
\label{eq:vformula}
\end{equation}
where $s$ depends only on the orientation of $d$ and on $(x\bmod4,\,y\bmod4)$:
\begin{equation}
\begin{aligned}
s_{\rm H}(x,y)&=-(-1)^{\floor{(x+1)/2}+\floor{(y+1)/2}},\\
s_{\rm V}(x,y)&=s_{\rm H}(x,y)\,(-1)^{\bar x\bar y}.
\end{aligned}
\label{eq:sfactor}
\end{equation}
\end{proposition}

Because $h(p_d)\bmod 4$ is fixed by position, the factor $\eps(p)\equiv(-1)^{\floor{h(p)/4}}$ records a single bit of the height field, namely which of the two allowed values of $h\bmod 8$ occurs at $p_d$. The vorticity pattern is therefore a fixed period-4 background multiplied by one bit of the height. The formula was found from all 32\,768 tilings of the $n=5$ Aztec diamond and is proven for every tiling of every region in Appendix~\ref{app:proofs}. Vorticities of individual tilings and their averages are shown in Appendix~\ref{app:figs}.

\subsection{No frustration and uniqueness}

Equation~\eqref{eq:vformula} satisfies every coupling of the island rule, including the corner couplings, in every tiling of every simply connected region (Lemma~\ref{lem:local} in Appendix~\ref{app:proofs}). The coupling network is thus never frustrated. Moreover, any two edge-adjacent dominoes share a perimeter island, so the coupling graph contains the adjacency graph of the tiling and is connected. The vorticity of each tetromino is therefore uniquely determined by the tiling, up to one global sign. The vorticity is not an extra degree of freedom on top of the dimers; it is a nonlocal functional of the dimer configuration, carried by the height field.

The global sign is not fixed by the boundary. Reversing all moments maps a ground state onto a ground state with the same unhappy vertices and the opposite vorticities, so each tiling corresponds to (at least) two ground states, and an average over both would give $\langle v\rangle=0$ identically. This Ising-like sign is chosen spontaneously, or by the preparation protocol, once for the whole sample. Throughout we fix this gauge by $v=+s\,\eps$ in Eq.~\eqref{eq:vformula}; all averages $\langle v\rangle$ and $\Psi$ refer to this convention. In an image the gauge is fixed by any fully ordered reference, such as a frozen corner (Sec.~\ref{sec:order}) or the sign of $V$ in samples with $V\neq0$.

\subsection{Conservation of the total vorticity}

A $2\times2$ flip of two parallel dominoes changes $h$ only at the center of the block, which is the reference vertex $p_d$ of no domino. By Eq.~\eqref{eq:vformula} only the two flipped tetrominoes change, and a direct evaluation shows that their sum does not (Appendix~\ref{app:proofs}): blocks whose lower-left cell has $x+y$ even carry opposite vorticities before and after the flip, blocks with $x+y$ odd carry equal ones of the same sign. Since all tilings of a simply connected region are connected by flips~\cite{Thurston1990}, we obtain the following.

\begin{theorem}[Topological total vorticity]
For a simply connected region, $V=\sum_d v(d)$ takes the same value for every ground state.
\end{theorem}

The value of $V$ can be read off any single tiling. For regions in which every row has even length, the all-horizontal (brick) tiling exists. In it the lattice line below each row is never crossed, so consecutive bricks share $\eps$ while $s_{\rm H}(x+2,y)=-s_{\rm H}(x,y)$; their vorticities cancel. This gives a closed boundary formula,
\begin{equation}
V=\sum_{\text{rows }j:\ \ell_j\equiv2\ (\mathrm{mod}\ 4)} s_{\rm H}(x_j,y_j)\,(-1)^{\floor{h(x_j,y_j)/4}},
\label{eq:Vboundary}
\end{equation}
where $\ell_j$ is the length of row $j$ and $(x_j,y_j)$ its lower-left vertex, which is a boundary vertex whose height is fixed by the sample shape. Table~\ref{tab:V} lists the resulting values. For rectangles $V\neq0$ requires a side $\equiv2\pmod4$. The theorem does not extend to regions with holes: a ring of tiles around a hole cannot be flipped, and the ground states split into winding sectors with different $V$ (Sec.~\ref{sec:holes}).

\begin{table}[b]
\caption{Total vorticity $V$ from Eq.~\eqref{eq:Vboundary}, valid for all sizes and confirmed by exhaustive enumeration ($n\le5$) and Monte Carlo sampling ($n,L\le60$). Nonzero values are $V=\pm2$, except for $W\times H$ rectangles with odd $H$, which have $V=\pm1$.}
\label{tab:V}
\begin{ruledtabular}
\begin{tabular}{lcc}
Sample & $V\neq0$ & $V=0$\\
\hline
Aztec, placement I & $n\equiv2\ (\mathrm{mod}\ 4)$ & otherwise\\
Aztec, placement II & $n\equiv1\ (\mathrm{mod}\ 4)$ & otherwise\\
$L\times L$, black corner & never & all $L$\\
$L\times L$, white corner & $L\equiv2\ (\mathrm{mod}\ 4)$ & $L\equiv0$\\
$W\times H$, $W$ even & see Eq.~\eqref{eq:rect} & $W\equiv0\ (\mathrm{mod}\ 4)$\\
\end{tabular}
\end{ruledtabular}
\end{table}

For a $W\times H$ rectangle with $W\equiv2\pmod4$, Eq.~\eqref{eq:Vboundary} reduces to
\begin{equation}
V=\sum_{j=0}^{H-1}c_{j\bmod4},\ \ c=\begin{cases}(-1,1,1,-1) & \text{black corner},\\ (1,1,-1,-1)& \text{white corner}\end{cases}.
\label{eq:rect}
\end{equation}

\section{Algebraic vorticity order}
\label{sec:order}

\subsection{The \texorpdfstring{$2\times2$}{2x2} superstructure}

\begin{figure*}[t]
\centering
\includegraphics[width=\textwidth]{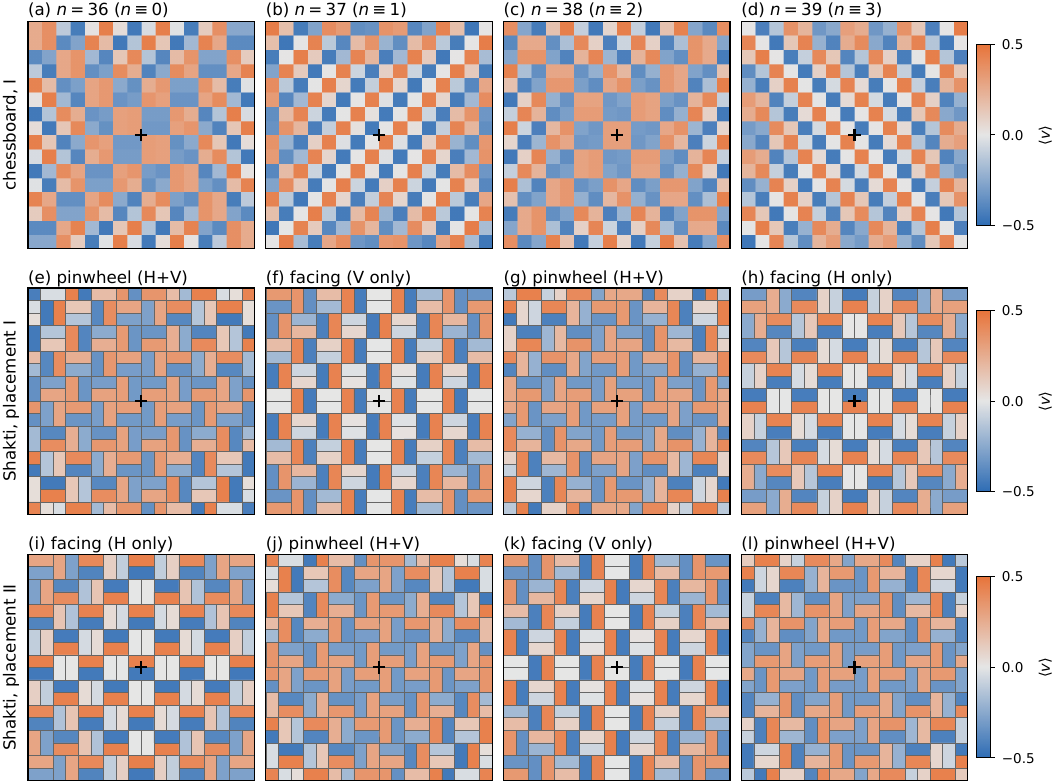}
\caption{The central vorticity pattern is selected by $n\bmod4$ and by the placement. Exact patterns at the center of the Aztec diamond for $n=36$--$39$, one size per residue class of $n\bmod4$, computed with the Kasteleyn method of Sec.~\ref{sec:exact} without sampling. (a)--(d) Chessboard view, placement I. (e)--(h) Shakti view of the same samples, in which each plaquette is colored by $\langle v\rangle$; (i)--(l) Shakti view for placement II. The cross marks the center. The $2\times2$ superstructure is present in every class. Its form is set by $\langle h(c)\rangle$ [Eq.~\eqref{eq:center}]: a pinwheel, in which horizontal (H) and vertical (V) plaquettes are ordered with equal weight, or a facing pattern, in which only one orientation is ordered. Placement I gives pinwheels for even $n$ and facing patterns for odd $n$; placement II gives the reverse.}
\label{fig:exactmaps}
\end{figure*}

In an individual ground state the vorticities look disordered (Appendix~\ref{app:figs}). Their average, however, is not zero. Inside the liquid region of the sample, $\langle v\rangle$ forms a checkerboard of $2\times2$ superblocks: in chessboard coordinates its Fourier weight sits at $(\pm\pi/2,\pm\pi/2)$ (Fig.~\ref{fig:exactmaps}). In the frozen regions outside the arctic circle of the Aztec diamond, $v$ is fully ordered: each frozen corner is a brick wall of parallel dominoes of a single type, and by Eq.~\eqref{eq:vformula} consecutive bricks carry opposite vorticities, so the corner shows a fixed pattern of alternating $v=\pm1$ with $|\langle v\rangle|=1$. The height has its maximal slope there, perpendicular to the bricks, so the phase of the pattern winds steadily across each frozen corner (Sec.~\ref{sec:phase}). The liquid order of the interior is the continuation of this frozen order across the arctic circle: the local amplitude drops from its saturated value at the circle to the algebraically small value of the center, which is the subject of this section.

The origin of the order is transparent from Eq.~\eqref{eq:vformula}. The factor $\eps=(-1)^{\floor{h/4}}$ is a period-8 function of $h$ whose leading harmonic is $e^{i\pi h/4}$. On large scales the dimer height is a Gaussian free field; with the present normalization, $h\to(4/\sqrt\pi)\,\phi$ with $\phi$ the Dirichlet Gaussian free field (GFF)~\cite{Kenyon2001,KenyonLectures}, so that $\mathrm{Var}\,h\simeq g\ln N$ with $g=8/\pi^2$. Hence
\begin{equation}
|\langle v\rangle|\propto\bigl|\langle e^{i\pi h/4}\rangle\bigr|\propto N^{-x},\qquad x=\frac{\pi^2 g}{32}=\frac14 .
\label{eq:exponent}
\end{equation}
The vorticity order is algebraic: an imprint of the boundary that penetrates the whole sample with a slowly decaying amplitude, as in a two-dimensional Coulomb phase~\cite{Henley2010,Nienhuis1987}.

\subsection{Exact results}
\label{sec:exact}

Height differences are signed counts of crossed dominoes, and every crossing changes $h$ by 4 relative to the dimer-free increment. Therefore $\eps(p)=\eps_0(p)(-1)^{N_\gamma(p)}$, where $N_\gamma$ is the number of dominoes crossing a lattice path $\gamma$ from a boundary vertex to $p$ and $\eps_0$ is fixed by the geometry of the path. The expectation is a ratio of Kasteleyn determinants~\cite{Kasteleyn1961,Temperley1961},
\begin{equation}
\langle\eps(p)\rangle=\eps_0(p)\,\frac{\det K_\gamma}{\det K},
\label{eq:seam}
\end{equation}
where $K_\gamma$ is the Kasteleyn matrix with the edges crossing $\gamma$ negated. Domino-resolved averages $\langle v(d)\rangle$ follow from cofactors (Appendix~\ref{app:kasteleyn}). The method reproduces full enumeration at $n=4$ to machine precision and yields exact rational numbers with fraction-free elimination.

Three exact statements follow. First, $\langle\eps(\text{center})\rangle$ vanishes identically for $n\equiv1,2\pmod4$ (placement I), and the values for $n=4m-1$ and $n=4m$ coincide; this is the exact form of the orientation rule of Fig.~\ref{fig:exactmaps}. Exact mod-4 structures of this kind are also known for the central placement probabilities of the Aztec diamond~\cite{Schoenfelder2025,Kim2026}. Second, the exact rationals have denominators $2^{2m(2m-1)}$ and numerators made of odd primes smaller than $4m$, which led us to the following closed form.

\begin{conjecture}[Central amplitude]
For the Aztec diamond of order $n=4m-1$ or $4m$ (placement I),
\begin{equation}
\bigl|\langle\eps(\text{center})\rangle\bigr|=2^{-2m(2m-1)}\prod_{\substack{j=1\\ j\ \text{odd}}}^{4m-1} j^{\,\min(4m-j,\;3j-4m)} .
\label{eq:product}
\end{equation}
\end{conjecture}

\begin{figure*}[t]
\centering
\includegraphics[width=\textwidth]{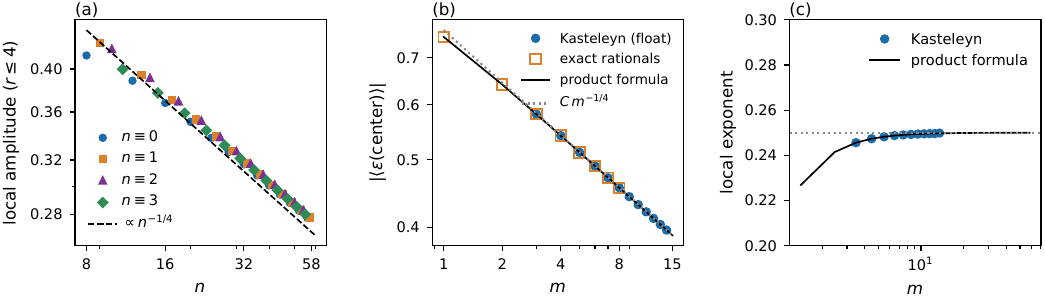}
\caption{The amplitude of the vorticity order decays as $n^{-1/4}$, and its central value follows the product formula. All data are exact (no statistical errors). (a) Local superstructure amplitude $A=[2(|\alpha|^2+|\beta|^2)]^{1/2}$ of $\langle v\rangle$, with $\alpha$ and $\beta$ the $(\pi/2,\pm\pi/2)$ Fourier components (Sec.~\ref{sec:phase}) over the cells within distance 4 of the center, versus $n$ for the four residue classes, with the $n^{-1/4}$ law (dashed). (b) $|\langle\eps(\text{center})\rangle|$ for $n=4m-1$: floating-point Kasteleyn values (circles), exact rationals (squares), the product formula of Eq.~\eqref{eq:product} (line), and its asymptote $C\,m^{-1/4}$ from Eq.~\eqref{eq:asym} (dotted). (c) Local exponent $-d\ln|\langle\eps\rangle|/d\ln m$, which approaches $1/4$.}
\label{fig:exactscaling}
\end{figure*}

Some exponents in Eq.~\eqref{eq:product} are negative ($j<4m/3$); the product is nevertheless always smaller than 1. Equation~\eqref{eq:product} reproduces all exact rationals for $m=1$--8 (e.g.\ $3/4$, $3\cdot5^3\cdot7/2^{12}$, $3^3\cdot5^3\cdot7^5\cdot11/2^{30}$) and the floating-point Kasteleyn values up to $m=14$ within their numerical error. Third, its asymptotic expansion follows from Stirling's formula and the hyperfactorial expansion,
\begin{equation}
\ln\bigl|\langle\eps(\text{center})\rangle\bigr|=-\tfrac14\ln m+\tfrac14+\tfrac13\ln2-3\ln A_{\rm G}+O(m^{-2}),
\label{eq:asym}
\end{equation}
with $A_{\rm G}$ the Glaisher--Kinkelin constant. The central amplitude is therefore $C\,m^{-1/4}$ with $C=e^{1/4}2^{1/3}A_{\rm G}^{-3}=0.767042\ldots$, in agreement with the extrapolated Kasteleyn data (Fig.~\ref{fig:exactscaling}). Using $\ln A_{\rm G}=\tfrac1{12}-\zeta'(-1)$, one has $C=2^{1/4}\,\bigl[2^{1/12}e^{3\zeta'(-1)}\bigr]$, where the bracket is the amplitude of the critical diagonal spin correlation of the square-lattice Ising model, $\langle\sigma_{0,0}\sigma_{N,N}\rangle\simeq2^{1/12}e^{3\zeta'(-1)}N^{-1/4}$~\cite{Forrester2018}. The two sequences are not identical (their subleading corrections differ), but the common constant is natural: in the exact bosonization of Dub\'edat~\cite{Dubedat2011} (extended to multiply connected domains in Ref.~\cite{Bayraktaroglu2025}), electric correlators of the dimer height such as $e^{i\pi h/4}$ are related to spin correlations of two independent Ising copies, and the exponent $\tfrac14=2\times\tfrac18$ is that of such a product. If Eq.~\eqref{eq:product} is proved---graphical condensation~\cite{Kuo2004} or symmetry factorization~\cite{Ciucu1997} are natural routes---the exponent $1/4$ at the sample center becomes a theorem that does not rely on the GFF limit.

\subsection{Shape dependence}

For a general region the amplitude at position $w$ follows from $\mathrm{Var}\,h(w)\simeq g\ln[N\rho_\Omega(w)]$, where $N$ is the linear size and $\rho_\Omega$ the conformal radius of the unit-scaled region $\Omega$:
\begin{equation}
A(w)\simeq c_\Omega\,[N\rho_\Omega(w)]^{-1/4}.
\label{eq:amplaw}
\end{equation}
For a $W\times H$ rectangle the central conformal radius is $H\rho(W/H)$ with $\rho(1)=0.539$ and $\rho(\infty)=2/\pi$. The short side therefore sets the amplitude, and elongation reduces it by at most 4\%. Monte Carlo simulations of 21 rectangles from $8\times8$ to $128\times32$ confirm this (Appendix~\ref{app:figs}); the global fit gives $A=0.82\,(H\rho)^{-0.24}$. Square boards have no frozen regions and a flat mean height; their amplitude $A$ is $0.38$ at $L=48$ (exponent $0.25\pm0.02$). Equation~\eqref{eq:amplaw} assumes a flat mean height and does not apply to the Aztec diamond, whose curved limit shape is treated in Sec.~\ref{sec:phase}.

\section{Boundary control: the vorticity phase field}
\label{sec:phase}

\subsection{One field behind all patterns}

At a vertex $p$ one has $h\equiv r(p)\pmod4$, so $e^{i\pi h/4}=e^{i\pi r(p)/4}\eps(p)$ exactly. The four corners of a cell carry all four residues, which allows us to define a local complex order parameter,
\begin{equation}
\Psi(c)=\frac12\sum_{p\in\partial c}e^{i\pi r(p)/4}\langle\eps(p)\rangle,
\label{eq:psi}
\end{equation}
with the inverse relation
\begin{equation}
\langle v(d)\rangle=s(d)\,\mathrm{Re}\bigl[e^{-i\pi r(p_d)/4}\,\Psi\bigr].
\label{eq:projection}
\end{equation}
All vorticity patterns---for any size, placement, or shape---are thus projections of one slowly varying field $\Psi\simeq A\,e^{i\varphi}$. The modulus $A$ is the amplitude of Sec.~\ref{sec:order}; the phase $\varphi$ selects which sublattice of plaquettes carries the pattern and with which sign. Writing the central pattern as $\langle v\rangle\simeq2\,\mathrm{Re}[\alpha e^{i\pi(x+y)/2}+\beta e^{i\pi(x-y)/2}]$, one finds $\alpha$ real and $\beta$ imaginary, so $\alpha$ lives on the horizontal and $\beta$ on the vertical plaquettes. The pair $(\alpha,i\beta)$ is a two-component vector whose angle follows the phase of $\Psi$: shifting the mean height by 2 rotates it by $\pi/2$ and moves the order from one plaquette orientation to the other.

\subsection{Phase map}

\begin{figure*}[t]
\centering
\includegraphics[width=\textwidth]{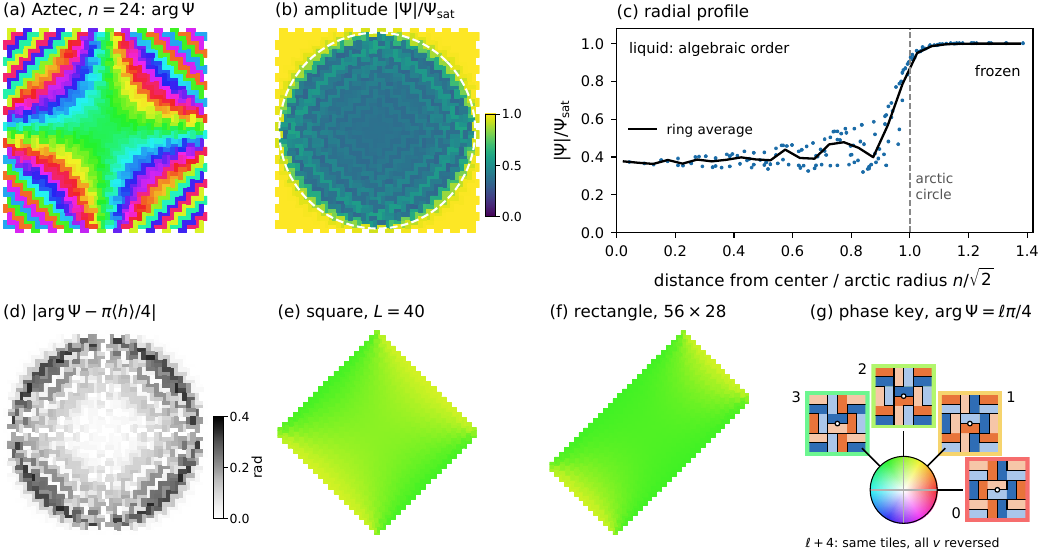}
\caption{The phase of $\Psi$ follows the mean height, $\arg\Psi\simeq\pi\langle h\rangle/4$, and its amplitude rises from the algebraic value at the center to complete order in the frozen corners. All maps are exact. (a) Exact $\arg\Psi$ for the Aztec diamond with $n=24$ (placement I), shown on the Shakti plaquettes. (b) Local amplitude $|\Psi|/\Psi_{\rm sat}$ of the same sample, where $\Psi_{\rm sat}=1.307$ is the value in the frozen corners (complete order); the dashed line is the arctic circle of radius $n/\sqrt2$ cells. (c) The amplitude of (b) versus distance from the center (dots: plaquettes; line: ring average). It is saturated in the frozen region, falls steeply inside the arctic circle, and levels off at $0.37$ at the center of this $n=24$ diamond, the algebraic amplitude of Sec.~\ref{sec:order}. (d) Absolute difference between $\arg\Psi$ and the prediction $\pi\langle h\rangle/4$ obtained from the exact mean height. (e), (f) Exact $\arg\Psi$ for a $40\times40$ chessboard square and a $56\times28$ chessboard rectangle (rotated by $45^\circ$ in the Shakti lattice), which have no frozen region; here the difference from $\pi\langle h\rangle/4$ is below 0.03\,rad everywhere (rms 0.006\,rad). (g) Phase key: the color wheel and the flat (basketweave) domino arrangements with $\arg\Psi=\ell\pi/4$, $\ell=0$--3, drawn as Shakti tetrominoes colored by vorticity ($v=+1$ orange, $v=-1$ blue; dark: strong arm, light: weak arm; the white dot is a fixed reference point). The sublattice of the parallel-domino $2\times2$ blocks sets $\ell$, and $\ell+4$ is the same arrangement with all vorticities reversed.}
\label{fig:phasemap}
\end{figure*}

For Gaussian fluctuations $\langle e^{i\pi h/4}\rangle=e^{i\pi\langle h\rangle/4}e^{-\pi^2\mathrm{Var}h/32}$, so the phase should be set by the mean height:
\begin{equation}
\arg\Psi(w)=\frac{\pi}{4}\,\langle h(w)\rangle .
\label{eq:phasemap}
\end{equation}
We tested Eq.~\eqref{eq:phasemap} against exact $\Psi$ with $\langle h\rangle$ computed exactly from the edge probabilities $K(b,w)K^{-1}(w,b)$ (Fig.~\ref{fig:phasemap}). On square and rectangular boards the rms phase error is $0.006$\,rad. In the Aztec diamond the relation is exact at the center, holds to better than $0.1$\,rad in the inner half of the liquid disk, and deviates by up to $0.35$\,rad only next to the arctic circle, where fluctuations are non-Gaussian. At large size $\langle h\rangle$ is the limit shape determined by the complex Burgers equation~\cite{Kenyon2007,KOS2006,Cohn1996}, so Eq.~\eqref{eq:phasemap} maps the variational limit shape directly onto the magnetic pattern.

Each phase has a simple real-space meaning. A flat basketweave tiling, in which $2\times2$ blocks of parallel dominoes alternate in orientation, has $\arg\Psi=\ell\pi/4$ exactly; the sublattice on which the blocks sit fixes $\ell\bmod4$, and a shift of the height by 4 reverses all vorticities and adds $\pi$ [Fig.~\ref{fig:phasemap}(g)]~\cite{Kim2026}. The phase of $\Psi$ thus tells which of these eight flat arrangements is locally favored. In the Aztec diamond the mean height is strongly curved, so the pattern rotates through all phases between the center and the frozen corners; in square boards the mean height is flat and the phase is uniform. The same contrast between the curved Aztec profile and the flat chessboard profile controls the sublattice structure of local $2\times2$ correlations, which is several times stronger in the Aztec diamond~\cite{Kim2026}.

\subsection{Symmetry-fixed central phase}
\label{sec:center}

If a region is invariant under a $90^\circ$ rotation $R$ about a vertex $c$, $R$ exchanges the chessboard colors and therefore maps $h\to c_0-h$. The uniform measure is invariant, which gives the exact result
\begin{equation}
\langle h(c)\rangle=\tfrac12\bigl[h(p)+h(Rp)\bigr]
\label{eq:center}
\end{equation}
for any boundary vertex $p$. The same argument pins the mean central height of the Aztec diamond used in the analysis of its local $2\times2$ correlations~\cite{Kim2026}. For the Aztec diamond, in the normalization $h=r$ at the bottom corner, this gives $\langle h(c)\rangle=3-n$ ($n$ even) and $1-n$ ($n$ odd) for placement I, and $n+2$ ($n$ even) and $n$ ($n$ odd) for placement II. We verified these integers with Kasteleyn calculations for $n=8$--40, and they agree with the exact central phases (Appendix~\ref{app:figs}). Since the symmetry maps $h-\langle h(c)\rangle\to-(h-\langle h(c)\rangle)$, $\langle e^{i\pi h(c)/4}\rangle$ is real up to the phase $e^{i\pi\langle h(c)\rangle/4}$, so Eq.~\eqref{eq:center} fixes $\arg\Psi(c)$ modulo $\pi$. Every mod-4 and placement rule then follows from $\langle h(c)\rangle$ (Table~\ref{tab:orient}): $\langle h(c)\rangle\equiv2\pmod4$ puts the order only on horizontal plaquettes, $\equiv0$ only on vertical plaquettes, and odd values on both with equal weight. The remaining sign is not fixed by symmetry; in all our Kasteleyn data it is the one given by $\langle h(c)\rangle$ modulo 8, as Eq.~\eqref{eq:phasemap} predicts. The size of the sample modulo 4 and its placement on the lattice are therefore the design parameters that select the imprinted pattern.

The two outcomes look different in real space [Fig.~\ref{fig:exactmaps}(e)--(l)]. For odd $\langle h(c)\rangle$ the phase is an odd multiple of $\pi/4$, so $|\alpha|=|\beta|$: both plaquette orientations are ordered with equal amplitude, and the ordered plaquettes around each $2\times2$ block turn like the vanes of a pinwheel. For even $\langle h(c)\rangle$ one of $\alpha,\beta$ vanishes: only horizontal ($\langle h(c)\rangle\equiv2$) or only vertical ($\equiv0$) plaquettes are ordered, and the ordered plaquettes form parallel pairs that face each other across unordered ones.

The parity of $n$ selects between the two in opposite ways for the two placements. A parity rule of the same origin appears in the exact local $2\times2$ probabilities of the Aztec diamond: the parity of $n$ decides which of the two interleaved sublattices of elementary squares dominates near the center, and within the dominant sublattice the preferred square follows the pinned mean central height~\cite{Kim2026}. At fixed $n$ the central heights of placements II and I differ by $(n+2)-(3-n)=2n-1$ for even $n$ and by $n-(1-n)=2n-1$ for odd $n$, an odd number in both cases. The one-cell shift that inverts the chessboard coloring therefore inverts the parity of $\langle h(c)\rangle$: placement I gives pinwheels for even $n$ and facing patterns for odd $n$, and placement II gives facing patterns for even $n$ and pinwheels for odd $n$. The exact central amplitudes confirm this; within distance 4 of the center, for example, placement I with $n=37$ gives $0.44$ on vertical and $0.03$ on horizontal plaquettes, whereas placement II with $n=37$ gives $0.31$ on both.

\begin{table}[b]
\caption{Which plaquettes carry the central order. The rule follows from $\langle h(c)\rangle\bmod4$ [Eq.~\eqref{eq:center}] and is confirmed exactly for the Aztec diamond and by Monte Carlo for square boards.}
\label{tab:orient}
\begin{ruledtabular}
\begin{tabular}{lccc}
$\langle h(c)\rangle\bmod4$ & 2 & 0 & odd\\
Ordered plaquettes & horizontal & vertical & both\\
\hline
Aztec, placement I & $n\equiv3$ & $n\equiv1$ & $n$ even\\
Aztec, placement II & $n\equiv0$ & $n\equiv2$ & $n$ odd\\
Square board & black corner & none & white corner\\
\end{tabular}
\end{ruledtabular}
\end{table}

\begin{figure*}[t]
\centering
\includegraphics[width=\textwidth]{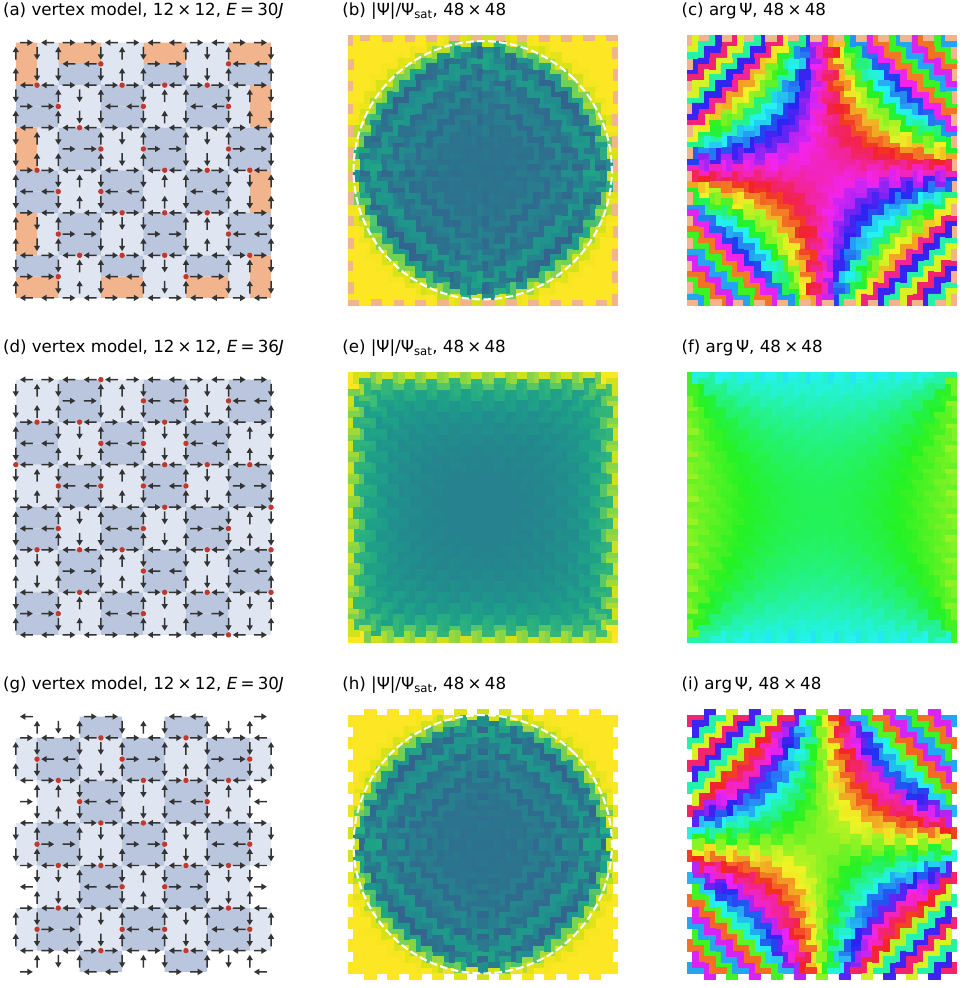}
\caption{Straight-edged arrays are Aztec diamonds: cut along block boundaries, their edge plaquettes close into free vortices and the tiled region is the placement-II Aztec diamond. Rows: (a)--(c) edges along block boundaries in the ground state, where the edge plaquettes are free and the tiled region is the placement-II Aztec diamond; (d)--(f) the same sample with every plaquette forced into a tetromino, for comparison; (g)--(i) edges through the middle of the blocks, where the tiled region is the placement-I Aztec diamond. (a), (d), (g) Exact states of the open vertex model for a $12\times12$ sample with their energies (arrows: island moments; red dots: unhappy vertices; blue: tiled plaquettes, dark horizontal and light vertical; orange: free edge plaquettes, each a complete vortex). In (a) and (g) the tiled region is the placement-II and placement-I Aztec diamond of order 5; forcing the edge plaquettes in (d) costs $(n+1)J=6J$. (b), (e), (h) Exact local amplitude $|\Psi|/\Psi_{\rm sat}$ and (c), (f), (i) exact $\arg\Psi$ for $48\times48$ samples, with the arctic circle (dashed) and the free edge plaquettes in orange; color scales as in Figs.~\ref{fig:phasemap}(b) and \ref{fig:phasemap}(g). With free edge vortices the sample is an Aztec diamond with frozen corners; with forced pairing it is liquid up to the edge.}
\label{fig:axis}
\end{figure*}

\subsection{Holes and pinned regions}
\label{sec:holes}

A central hole adds a second boundary whose rim vorticity is fixed up to a winding sector $k$ (a rim height offset $4k$). Kasteleyn calculations with a flux through the hole give the sector weights. For holes of size $H\equiv0\pmod4$ one sector dominates overwhelmingly (99.1\% for $H=8$ in an $L=48$ board), so the hole pins the pattern by itself. For $H\equiv2\pmod4$ two sectors have equal weight, and the rim vorticity is a free Ising variable that an external pinning field can select.

When the rim pattern agrees with the one imposed by the outer boundary, the order is reinforced: the local amplitude reaches $0.61$ at the rim, compared with $0.38$ in the same $L=48$ board without a hole (Appendix~\ref{app:figs}). When the rim pattern is opposite, the order does not cancel. Instead $\Psi$ rotates by $\pi$ across the annulus while $|\Psi|$ stays unchanged, because the forcing moves only the mean height, not its fluctuations. Halfway through the twist the order is carried entirely by the other plaquette orientation. The twist follows the logarithmic interpolation of $\langle h\rangle$: the crossover ring is predicted at the geometric mean of the hole and sample radii ($9.8$ cells) and observed at $9$ cells. A pinned structure thus reinforces the imprinted pattern when aligned and produces a horizontal/vertical crossover ring when misaligned.

\subsection{Straight-edged samples}
\label{sec:axis}

Fabricated Shakti arrays usually have straight edges along the lattice axes. By the $45^\circ$ rotation of Sec.~\ref{sec:mapping} such a sample is a chessboard diamond. Consider an $X\times X$ sample with even $X$. If its edges run along the boundaries of the $2\times2$ plaquette blocks, as for an array cut from the lattice in whole blocks, its complete plaquettes form the placement-II Aztec diamond of order $n=X/2-1$ together with $2(n+1)$ plaquettes along the edge. These edge plaquettes do not take part in the tiling. Each of them contains no unhappy vertex, and all of its islands circulate in one sense, so it forms a complete vortex on its own [Fig.~\ref{fig:axis}(a), orange]. Exact ground states of the open vertex model confirm this for $n=5$ and 6 (integer programming; unhappy vertex $J$, type-II and charged vertices $3J$): the energy is $n(n+1)J$, one unhappy vertex per domino of the Aztec diamond, and every edge plaquette carries a closed circulation. If instead the edges run through the middle of the blocks, the complete plaquettes form exactly the placement-I Aztec diamond, and the islands of the cut plaquettes are left at the edge [Fig.~\ref{fig:axis}(g)].

The free edge vortices decide the outcome. If every plaquette of the block-aligned sample is instead forced into a tetromino, the chessboard region is a diamond bounded by staircases with steps of two cells. Along such a staircase the dimer-free height zigzags within a range of 4, whereas along the unit staircase of the Aztec diamond it rises monotonically (range 46 for $n=23$). The forced boundary is therefore flat, and the sample behaves like a chessboard square: it is liquid up to the edge, with a nearly uniform phase, no frozen corners, and a central $|\Psi|/\Psi_{\rm sat}$ of 0.44 instead of 0.38 for the $48\times48$ sample [Figs.~\ref{fig:axis}(d)--\ref{fig:axis}(f)]. This state is not the ground state. Pairing the $2(n+1)$ edge plaquettes adds $n+1$ unhappy vertices, and the open vertex model gives exactly $E_{\rm forced}-E_{\rm gs}=(n+1)J$ for $n=5$ and 6 [Figs.~\ref{fig:axis}(a) and \ref{fig:axis}(d)]. Because of this cost the edge plaquettes close into complete vortices in the ground state, and the Aztec boundary with its frozen corners takes over. The same straight edge thus imposes either a flat or a maximally tilted boundary height, and at low temperature the energetics selects the tilted one.

At finite temperature the boundary can be excited in two ways. Edge plaquettes can pair: since every tetromino carries one unhappy vertex, the open vertex model gives the energy $J[n(n+1)+|S|/2]$ when a set $S$ of edge plaquettes is paired, which we checked for random sets at $n=5$. In addition, a boundary vertex of coordination 3 can become unhappy while covering a single plaquette. This half dimer also costs $J/2$, such sites occur along every edge, and each has two spin states. Both excitations are monomers on the outer face, the free edge plaquettes with fugacity $e^{J/2T}$ and the half-dimer sites with fugacity $2e^{-J/2T}$ relative to a dimer. The partition function is therefore a Pfaffian, which yields exact results at all temperatures (Appendix~\ref{app:sim}). This boundary model reproduces exactly the first excited level of the full vertex model, which we obtained from its exact density of states for $n=3$ and 5. Each boundary defect, a paired edge plaquette or an occupied half-dimer site, shifts the boundary height beyond it by 4, so it acts as a half vortex at the edge and reverses the sign of $\Psi$ in the adjacent sector. At low temperature their mean number is $\langle N_d\rangle=2^{n+4}e^{-J/T}$, half of them paired edge plaquettes and half occupied half-dimer sites [Fig.~\ref{fig:edgeT}(a)]. The imprinted order is therefore lost once $\langle N_d\rangle\approx1$, at
\begin{equation}
T_{\rm on}=\frac{J}{(n+4)\ln2},
\label{eq:Ton}
\end{equation}
about $0.05J$ for a $48\times48$ array. At $T_{\rm on}$ the central amplitude has dropped by about a quarter, and at $1.25\,T_{\rm on}$ to below half [Fig.~\ref{fig:edgeT}(c)]. Above $T_{\rm on}$ the boundary does not become flat. Even at infinite temperature only about 45\% of the edge plaquettes are paired, and the randomly placed boundary defects leave a residual central amplitude that decreases with size, from 0.08 for $n=11$ to 0.02 for $n=47$. The flat liquid state of Figs.~\ref{fig:axis}(d)--\ref{fig:axis}(f) is thus never reached thermally.

Below $T_{\rm on}$ a straight-edged array is therefore an Aztec diamond in either case. It has frozen corners and an arctic circle, and its phase rotates toward the corners [Figs.~\ref{fig:axis}(b), \ref{fig:axis}(c), \ref{fig:axis}(h), and \ref{fig:axis}(i)]. The termination selects the placement and, with it, the central pattern of Sec.~\ref{sec:center}: an array cut along block boundaries (placement II) has a pinwheel at its center when $n$ is odd ($X\equiv0\pmod4$) and a facing pattern when $n$ is even ($X\equiv2\pmod4$), and an array cut through the blocks has the opposite. Arrays cut through the blocks have no edge plaquettes, but their half-dimer sites, on every other boundary cell, give $\langle N_d\rangle=2^{n+2}e^{-J/T}$ and hence the onset $J/[(n+2)\ln2]$, slightly above the block-aligned value for the same $n$. In units of its own onset temperature each termination loses the imprint along nearly the same curve [Fig.~\ref{fig:edgeT}(c)]. Since the onset decreases as $1/n$, large arrays must be equilibrated far below the unhappy-vertex energy, whatever their termination.

\begin{figure*}[t]
\centering
\includegraphics[width=\textwidth]{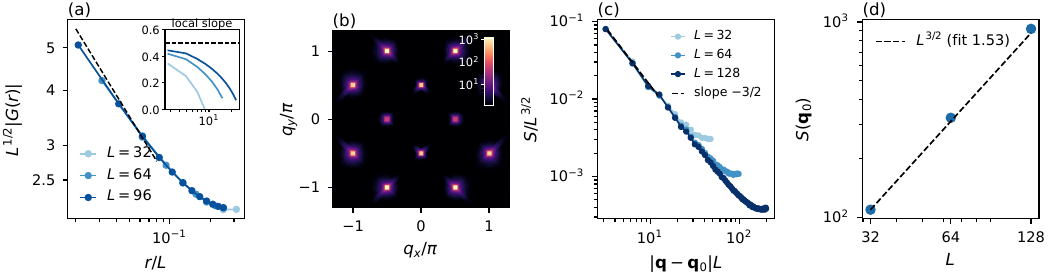}
\caption{The vorticity order has algebraic correlations and power-law superlattice peaks with the predicted exponents. (a) Exact phase correlation on square boards: $L^{1/2}|G(r)|$ versus $r/L$ collapses for $L=32$--96; inset: the local slope approaches $1/2$. (b) Vorticity structure factor for $L=128$ (Monte Carlo, logarithmic color scale; peaks are widened by a $7$-pixel maximum filter for visibility). (c) Scaled peak profile $S/L^{3/2}$ versus $|\bm q-\bm q_0|L$ with the $-3/2$ tail. (d) Peak height versus $L$, with fitted slope 1.53.}
\label{fig:corr}
\end{figure*}

\section{Correlations and scattering signature}
\label{sec:corr}

The two-point function of the vorticity phase can also be computed exactly. With a seam running from $p$ to $q$,
\begin{equation}
G(p,q)=\bigl\langle e^{i\pi[h(q)-h(p)]/4}\bigr\rangle=e^{i\pi\Delta h_0/4}\,\frac{\det K_\gamma}{\det K},
\label{eq:twopoint}
\end{equation}
where $\Delta h_0$ is the dimer-free height increment along the seam. On square boards of size $L=32$, 64, and 96, $L^{1/2}|G(r)|$ collapses onto a single function of $r/L$ to within 1\% [Fig.~\ref{fig:corr}(a)]. This confirms the two-point exponent $2x=1/2$ without adjustable parameters. The collapse function is not the pure Dirichlet-GFF prediction (deviations of up to 15\%), reflecting the non-Temperleyan corners of an even board~\cite{Russkikh2018}; the exponent is unaffected.

In an experiment the natural observable is the structure factor. We computed $S(\bm q)=L^{-2}\langle|\sum_c v(c)e^{i\bm q\cdot\bm R_c}|^2\rangle$ for the vorticity field on the plaquette centers $\bm R_c$ (Shakti units, lattice constant $a=1$) with Monte Carlo for $L=32$--128 [Fig.~\ref{fig:corr}(b)--(d)]. Peaks appear at $\bm q_0=\pi(\pm\tfrac12,\pm1)$, $\pi(\pm1,\pm\tfrac12)$, $\pi(0,\pm\tfrac12)$, and weakly at $\pi(\pm\tfrac12,0)$, with relative weights set by the boundary. Algebraic order with exponent $x$ gives
\begin{equation}
S(\bm q_0+\bm\kappa)=L^{2-2x}\,F(\kappa L),\qquad F(u)\sim u^{-(2-2x)}\ (u\gg1),
\label{eq:sq}
\end{equation}
that is, peak heights $\propto L^{3/2}$ and tails $\propto|\bm q-\bm q_0|^{-3/2}$. A fit to the three sizes $L=32$, 64, and 128 gives a peak-height exponent of $1.53$ and a scaling collapse with the $-3/2$ tail for $\kappa L\lesssim12$. Such power-law superlattice peaks distinguish the vorticity order from true long-range order ($S\propto L^2$) and from short-range order ($S\propto L^0$).

\begin{figure*}[t]
\centering
\includegraphics[width=\textwidth]{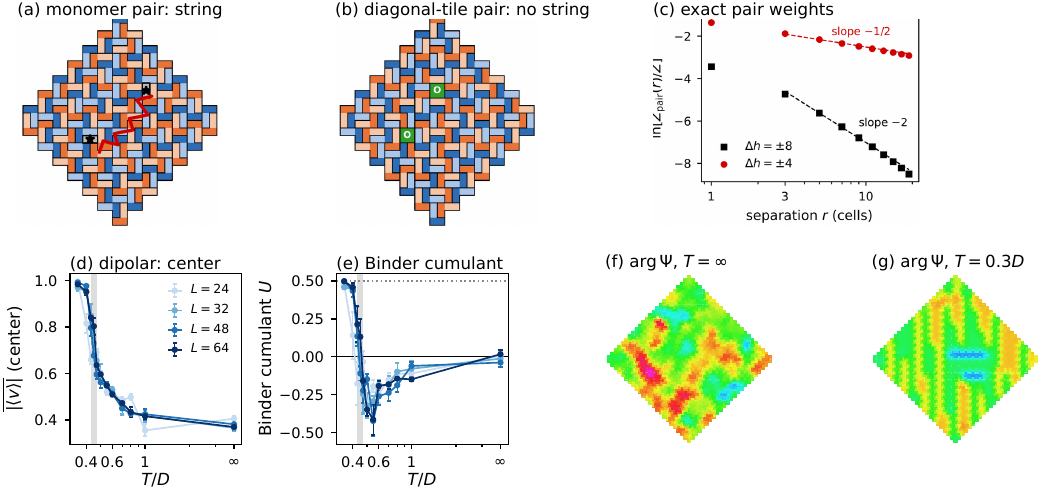}
\caption{Half vortices are confined by strings while full vortices are free, and dipolar interactions keep the algebraic order down to $T_c\approx0.45D$. (a), (b) Single tilings of a $14\times14$ board with vorticities from the island rule; red segments mark a minimal set of violated couplings. (a) Two monomers (stars, half vortices) are joined by a string. (b) Two diagonal tiles (full vortices) cause no violated coupling. (c) Exact pair weights on a $64\times64$ board: $r^{-1/2}$ for half vortices ($\Delta h=\pm4$) and $r^{-2}$ for full vortices ($\Delta h=\pm8$). (d)--(g) Dipolar interactions within the ground-state manifold (point dipoles, $L\times L$ boards, $L=24$, 32, 48, and 64 from light to dark). (d) Mean $|\langle v\rangle|$ in the central window and (e) Binder cumulant of the crystal order parameter versus temperature; points are means over four independent replicas, error bars standard errors, and the gray band marks $T_c=(0.45\pm0.02)D$. (f), (g) Phase field of single states at $L=48$ (colors as in Fig.~\ref{fig:phasemap}); at $T=0.3D$ the crystal has formed, with domains of different locked heights.}
\label{fig:robust}
\end{figure*}

\section{Robustness and read-out}
\label{sec:robust}\label{sec:defects}\label{sec:dipolar}

The vorticity order of Secs.~\ref{sec:height}--\ref{sec:phase} lives in the degenerate ground-state manifold. Real arrays contain defects, and the dipolar interaction lifts the degeneracy. Here we summarize how the order responds and how it can be read out; technical details are given in Appendix~\ref{app:sim}.

\emph{Defects.} A point defect is characterized by its winding: going once around it, $h$ changes by $\Delta h$ and the phase $\pi h/4$ by $\pi\Delta h/4$. Since $\eps=(-1)^{\floor{h/4}}$ has period 8 in $h$, two classes arise. Half vortices ($\Delta h=\pm4$), such as a monomer (an uncovered plaquette), reverse the sign of $\Psi$. Monomers are the topological charges identified in the kinetics of Shakti arrays by Lao \emph{et al.}~\cite{Lao2018}. Every loop of tetrominoes around a half vortex then contains a violated coupling, so a string of violated couplings joins it to another half vortex or to the edge [Fig.~\ref{fig:robust}(a); in exact maps of $\Psi$ the string is the line where opposite hues meet (Appendix~\ref{app:figs})]. Its length, and its energy in the Shakti vertex model, grow linearly with the separation. The logarithmic monomer interaction of the plain dimer model, $Z_{\rm pair}\propto r^{-1/2}$~\cite{Fisher1963}, thus becomes linear confinement, and half vortices can only bind in pairs or leave through the edge. Full vortices ($\Delta h=\pm8$), such as the diagonal tiles, rotate $\Psi$ by $\pm2\pi$ and leave no string [Fig.~\ref{fig:robust}(b)]. Their exact pair weight decays as $r^{-2x_v}$ with $x_v=1$ [Fig.~\ref{fig:robust}(c)], four times the half-vortex value, as expected for a doubled charge in a Coulomb gas~\cite{Nienhuis1987}. Since $x_v<2$, a finite density of free full vortices is relevant~\cite{Kosterlitz1973,Jose1977}: it screens the algebraic order beyond a length $\xi\sim\lambda_{\rm d}^{-1/(2-x_v)}=\lambda_{\rm d}^{-1}$, where $\lambda_{\rm d}$ is the defect fugacity.

\emph{Dipolar interactions.} Every tiling fixes all island moments through the templates, so the dipolar energy is a function on the ground-state manifold. The cheapest excitation out of the manifold costs at least $3.8D$ for point dipoles, $D=\mu_0\mu^2/4\pi a^3$ ($\mu$: island moment), which makes the ordering two-stage, as in dipolar kagome ice~\cite{Chern2011}. Metropolis height flips on $L\times L$ boards ($L=24$--64, four independent replicas each) give $2x=0.50\pm0.03$ at $T=\infty$ and $2x=0.58\pm0.04$ for $0.55D\le T\le D$. The order thus stays algebraic, with a slightly larger exponent, as expected for interacting dimers whose Coulomb-gas coupling varies continuously~\cite{Alet2005}. The boundary-control results of Sec.~\ref{sec:phase} depend only on the mean height and carry over [Figs.~\ref{fig:robust}(d) and \ref{fig:robust}(f)]. The algebraic window is $0.55D\lesssim T\ll3.8D$, less than a decade.

At $T_c=(0.45\pm0.02)D$, where the Binder cumulants of the crystal order parameter cross, the height locks into a flat crystal with wave vector $(\pi/2,\pi/2)$ in chessboard coordinates. Every tetromino then has a fixed vorticity, $|\langle v\rangle|=1$, so the algebraic order turns into true long-range order. A negative minimum of the Binder cumulant above $T_c$ [Fig.~\ref{fig:robust}(e)] suggests that the transition is weakly first order (Appendix~\ref{app:sim}). A boundary-selected phase [Eq.~\eqref{eq:center}] that is incompatible with the crystal leaves grain boundaries [Fig.~\ref{fig:robust}(g)]. Dumbbell moments give the same crystal at a similar $T_c$, in line with interacting-dimer models whose height locks at low temperature~\cite{Alet2005}. Chern \emph{et al.} found that dipolar interactions select a staggered arrangement of vertex charges with many nearly degenerate layered states~\cite{Chern2013}; whether our locked crystal and its degenerate variants coincide with those states remains to be checked.

\emph{Equilibration.} The boundary imprint is a property of the equilibrium ensemble. Observing it requires the height field to relax on the scale of the sample, which takes of order $L^2$ flip times multiplied by the activation barrier of a flip, and Shakti kinetics are known to be topologically constrained~\cite{Lao2018,Priyanka2025}, and real-space imaging of thermally active arrays has resolved the crossover from ergodic to arrested dynamics~\cite{Saccone2023}. Thermally active arrays annealed slowly through the vertex gap are therefore the natural candidates; field-demagnetized arrays need not reach the imprinted state.

\emph{Read-out.} Each magnetic image can be demodulated into $\Psi$ via Eq.~\eqref{eq:psi}, after fixing the global sign of Sec.~\ref{sec:height} image by image. Uniform color signals order, a point around which the color runs once through the color wheel marks a full vortex, and a line across which the color jumps marks the string of a half vortex. In reciprocal space the order gives the power-law superlattice peaks of Sec.~\ref{sec:corr}, with heights $\propto L^{3/2}$. Resolving $\langle v\rangle\approx0.4$ to 10\% on a single plaquette requires about 500 independent images, fewer when symmetry-equivalent plaquettes are averaged.

\section{Conclusions}
\label{sec:conclusion}

We have shown that the degenerate ground states of Shakti artificial spin ice host a hidden collective order: the vorticity of the L-tetrominoes formed by unhappy vertices and their plaquettes. Far from being an independent Ising variable, this vorticity is a nonlocal functional of the dimer configuration, $v=s(-1)^{\floor{h/4}}$. Three physical consequences stand out. First, the total vorticity is a topological invariant fixed by the sample boundary, for which we derived a closed formula. Second, the vorticity displays algebraic order with exponent $1/4$, imprinted by the boundary into the bulk; its amplitude at the Aztec center appears to follow a closed product formula with the constant $e^{1/4}2^{1/3}A_{\rm G}^{-3}$, which links it to the critical Ising model. Third, a single phase field $\Psi\simeq e^{i\pi\langle h\rangle/4}$, set by the dimer limit shape, organizes all patterns, so that sample size, placement, holes, and pinning become knobs that select and rotate the magnetic texture. Point defects fall into two topological classes: half vortices are confined by strings, while full vortices are free and screen the order. In our simulations dipolar interactions preserve the algebraic phase down to $T_c\approx0.45D$ and then lock it into long-range order.

For experiments the results reduce to a few rules. The order fills the liquid region: chessboard squares and rectangles, rotated by $45^\circ$ in the Shakti lattice, are liquid throughout, whereas arrays with straight edges along the Shakti axes are Aztec diamonds with frozen corners, realizing placement II when cut along block boundaries, provided they equilibrate below $T_{\rm on}=J/[(n+4)\ln2]$ [Eq.~\eqref{eq:Ton}]; arrays cut through the blocks obey the same limit with $n+2$ in place of $n+4$. The central pattern follows from $\langle h(c)\rangle$ (Table~\ref{tab:orient}), so sample size modulo 4 and edge termination select a pinwheel or a facing pattern, and samples with $V=\pm2$ carry a net circulation. In reciprocal space the order appears as power-law superlattice peaks with heights $\propto L^{3/2}$. The amplitude decays only as $(\text{size})^{-1/4}$; at the center of an Aztec diamond with $n\approx40$ it is still about one third of complete order ($|\Psi|/\Psi_{\rm sat}\approx0.33$); among samples with a flat mean height, compact shapes with a large conformal radius maximize it. Holes of size $\equiv0\pmod4$ pin the pattern, holes of size $\equiv2\pmod4$ act as switchable Ising elements, and misaligned pinning creates a horizontal/vertical crossover ring. Algebraic order is expected for $T_c\approx0.45D<T\ll4$--$8D$, the vertex gap, and every image must be referred to a common global sign.

These results turn the abstract height field of the dimer model into a directly observable magnetic texture and offer a new way to use boundaries as a design tool in artificial spin ice. Several questions remain open: a proof of the product formula~\eqref{eq:product}, possibly through the Ising correspondence; the universality class and variants of the dipolar locking transition, including realistic island shapes and kinetics; and whether analogous vorticity orders exist in other vertex-frustrated geometries. For Tetris ice, whose ground states reduce to one-dimensional structures~\cite{Gilbert2016}, and Santa Fe ice, whose unhappy vertices form strings~\cite{Zhang2021}, we expect at most short-range vorticity correlations, which singles out Shakti as the geometry with a two-dimensional height field. The recent observation of memory effects from topological constraints in Shakti ice~\cite{Priyanka2025} suggests that the vorticity may also be a useful variable for nonequilibrium studies.

\begin{acknowledgments}
This research was supported by the National Research Foundation (NRF) of Korea funded by the Korean Government (NRF-2023R1A2C1006050). The authors used Claude (Anthropic) to assist with writing code for the numerical computations and figures and with editing the manuscript; all results were checked by the authors, who take full responsibility for the content.
\end{acknowledgments}

\section*{Data availability}
The computer codes that support the findings of this article, and the scripts that generate all data and figures, are openly available at \url{https://github.com/cywon-commits/shakti-vorticity}.

\appendix

\section{Proofs}
\label{app:proofs}

\emph{The map.} Shakti unit squares are $[p,p+1]\times[q,q+1]$, and the block $(a,b)=[2a,2a+2]\times[2b,2b+2]$ is horizontal if $a+b$ is even. The cells of one fundamental domain map as follows: $(0,0)\mapsto$ squares $(0,1),(1,1)$ [H]; $(1,0)\mapsto(1,2),(1,3)$ [V]; $(1,1)\mapsto(0,4),(1,4)$ [H]; $(2,0)\mapsto(2,3),(3,3)$ [H]; $(1,-1)\mapsto(2,2),(3,2)$ [H]; $(2,1)\mapsto(2,4),(2,5)$ [V]; $(2,-1)\mapsto(4,2),(4,3)$ [V]; $(3,0)\mapsto(3,4),(3,5)$ [V]. All other cells follow by the translations $(2,2)\mapsto(0,4)$ and $(2,-2)\mapsto(4,0)$.

\emph{Island rule.} A shared island lies on the common boundary of plaquettes that belong to different tetrominoes. A counterclockwise circulation of one traverses that edge opposite to a counterclockwise circulation of the other. Agreement of the island orientation therefore requires opposite vorticities, with one sign change if the island is the defect island of exactly one of the tiles.

\begin{lemma}[Locality]\label{lem:local}
In any tiling of any simply connected region, Eq.~\eqref{eq:vformula} satisfies every island coupling.
\end{lemma}
\begin{proof}
Every coupled pair of dominoes $a,b$ shares a chessboard vertex. Edges on the perimeter of a domino are never crossed by a domino, so $h$ can be carried from $p_a$ along the perimeter of $a$ to the shared vertex and along the perimeter of $b$ to $p_b$. The difference $h(p_b)-h(p_a)$ is therefore a dimer-free increment that depends only on the relative placement of $a$ and $b$. The coupling sign also depends only on that placement. Shifting $h$ by 4 flips both $\eps$ factors, so $v(a)v(b)$ depends only on the relative placement and the position modulo 4. All 704 coupled placements (both orientations, 16 positions) were evaluated and satisfy the rule. Since this list is complete, the check is a proof.
\end{proof}

\emph{Connectivity.} The template fixes all ten perimeter islands of a tetromino. Two chess-adjacent cells map to plaquettes that share exactly one unit edge (verified on one period of the map); that edge separates different plaquettes, so it is an island. Edge-adjacent dominoes therefore share a prescribed island and are coupled. In a connected region any two dominoes are joined by a chain of edge-adjacent dominoes, so the coupling graph is connected and $v$ is unique up to a global sign.

\begin{figure}[t]
\centering
\includegraphics[width=\columnwidth]{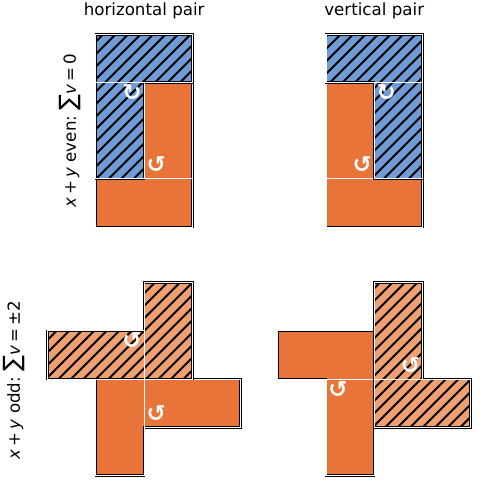}
\caption{A $2\times2$ flip leaves the summed vorticity of the two flipped tetrominoes unchanged. Upper row: blocks whose lower-left cell has $x+y$ even carry opposite vorticities in both fillings ($\sum v=0$). Lower row: blocks with $x+y$ odd carry equal vorticities of the same sign in both fillings ($\sum v=\pm2$). Hatching distinguishes the two tetrominoes; one representative sign choice is shown.}
\label{fig:flip}
\end{figure}

\emph{Flip invariance.} A flip at block center $c$ changes only $h(c)$, by $\pm4$, and $c$ is the reference vertex of no domino. The flipped dominoes use the reference vertices $a=c-(1,1)$ together with $b=c-(1,0)$ (horizontal pair) or $d=c-(0,1)$ (vertical pair). The left and bottom block edges are never crossed, so $h(b)$ and $h(d)$ follow from $h(a)$ by one step. The block vorticity is thus a function of $(c\bmod4,\ h(a)\bmod8)$; evaluating Eqs.~\eqref{eq:vformula}--\eqref{eq:sfactor} for all 16 positions $c\bmod4$ shows that it is unchanged by the flip (Fig.~\ref{fig:flip}); the second value of $h(a)\bmod8$ reverses all vorticities and changes nothing.

\emph{Boundary formula.} In the brick tiling the lattice line under each row is never crossed, so $h(x+2,y)=h(x,y)$ along it, while $s_{\rm H}(x+2,y)=-s_{\rm H}(x,y)$ by Eq.~\eqref{eq:sfactor}. Consecutive bricks cancel, rows of length $\equiv0\pmod4$ contribute nothing, and rows of length $\equiv2\pmod 4$ contribute the vorticity of their first brick, which gives Eq.~\eqref{eq:Vboundary}.

\emph{Center height.} The $90^\circ$ rotation $R$ about $c$ maps cell $(i,j)$ to a cell of the opposite color, so the height of the rotated tiling is $c_0-h\circ R^{-1}$. Invariance of the uniform measure gives $\langle h(c)\rangle=c_0-\langle h(c)\rangle$, and $c_0=h(p)+h(Rp)$ for any boundary vertex $p$.

\begin{figure*}[tp]
\centering
\includegraphics[width=\textwidth]{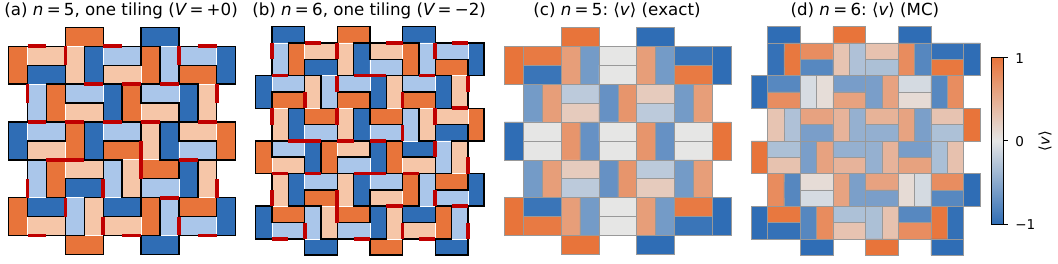}
\caption{Individual ground states have disordered vorticities, but their average forms a $2\times2$ superstructure (Aztec diamond, placement I). (a), (b) One tiling each for $n=5$ and $n=6$, with $v$ obtained from the island rule; orange (blue) is $v=+1$ ($-1$), dark (light) shading the strong (weak) arm, and red bars mark the defect islands. The total vorticity $V$ given in brackets is the same for all tilings of a sample. (c) Average $\langle v\rangle$ per plaquette, exact over all 32\,768 tilings for $n=5$. (d) The same for $n=6$ from $2\times10^4$ independent samples (standard error at most $0.007$ per plaquette).}
\label{fig:examples}
\end{figure*}

\begin{figure*}[tp]
\centering
\includegraphics[width=0.55\textwidth]{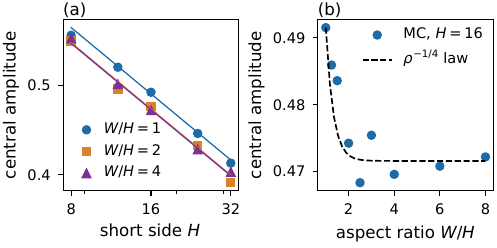}
\caption{The amplitude is set by the conformal radius: the short side dominates and elongation lowers it by at most 4\% (Monte Carlo, black corner). The amplitude $A$ is defined as in Fig.~\ref{fig:exactscaling}(a), over the cells within $H/4$ of the center; the scatter of the points in (b), about 0.005, indicates the statistical uncertainty. (a) $A$ versus the short side $H$ for aspect ratios $W/H=1$, 2, and 4; lines are the global fit $A=0.82\,(H\rho)^{-0.24}$. (b) Dependence on the aspect ratio at $H=16$, compared with the conformal-radius law of Eq.~\eqref{eq:amplaw}.}
\label{fig:rect}
\end{figure*}

\section{Supplementary figures}
\label{app:figs}

\begin{figure*}[tp]
\centering
\includegraphics[width=\textwidth]{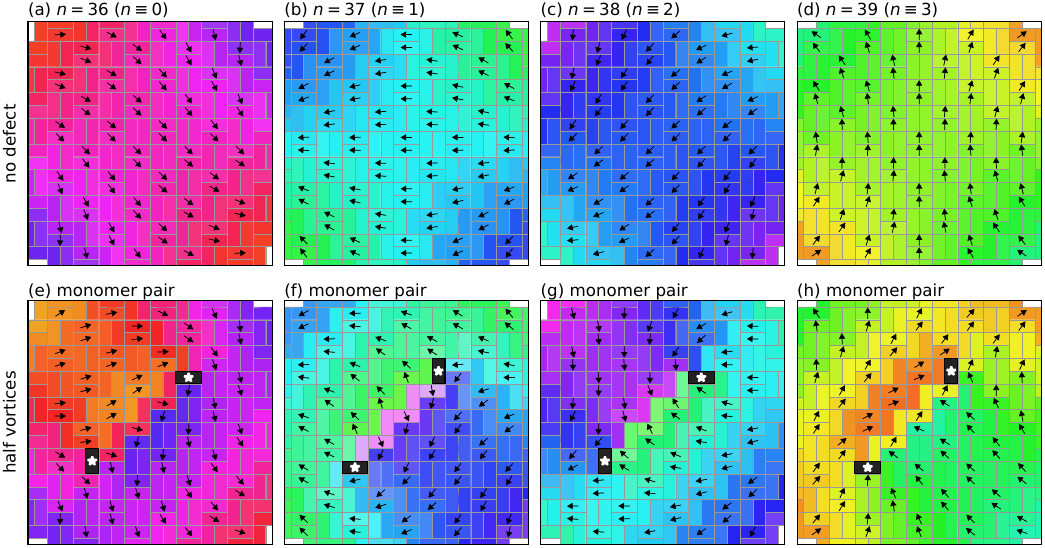}
\caption{The central phase is uniform and set by $n\bmod4$, and a pair of monomers makes it wind by $\pi$ around each. Exact $\Psi$ at the center of the Aztec diamond (placement I, $n=36$--39), shown on the Shakti plaquettes; hue is $\arg\Psi$ (colors as in Fig.~\ref{fig:phasemap}), brightness $|\Psi|$ (from 0 to 0.45), and arrows give the direction of $\Psi$. (a)--(d) Without defects the phase is uniform and changes with $n\bmod4$ as predicted by Eq.~\eqref{eq:center}. (e)--(h) With a pair of monomers (stars) seven cells apart, the phase winds by $\pi$ around each monomer and opposite hues meet along a line, the string (Sec.~\ref{sec:defects}).}
\label{fig:psi}
\end{figure*}

\begin{figure*}[tp]
\centering
\includegraphics[width=\textwidth]{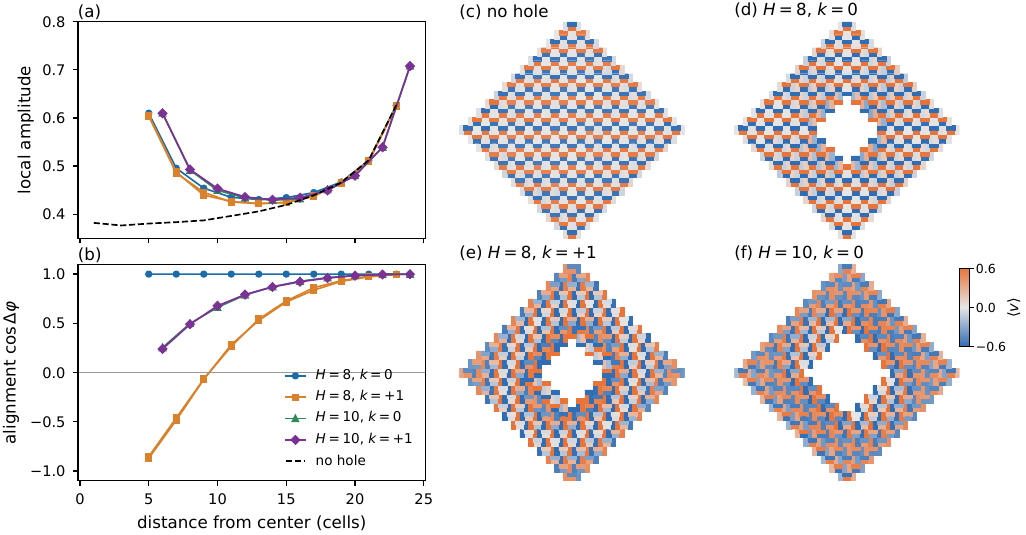}
\caption{A hole pins the pattern when its rim agrees with the outer boundary and twists it by $\pi$ through a crossover ring when it disagrees. Central holes in an $L=48$ chessboard square, rotated by $45^\circ$ in the Shakti lattice; Monte Carlo in fixed winding sectors $k$. For $H=10$ the two sectors are degenerate and their curves coincide. (a) Local superstructure amplitude and (b) alignment $\cos\Delta\varphi$ with the hole-free pattern versus distance from the center, for holes of size $H=8$ and 10. (c)--(f) $\langle v\rangle$ maps. In the natural sector ($k=0$) of the $H=8$ hole the pattern is reinforced; in the forced sector ($k=+1$) it twists by $\pi$ through a ring in which the other plaquette orientation carries the order. For $H=10$ the two sectors are degenerate.}
\label{fig:holes}
\end{figure*}

The figures of this appendix collect results that support the main text but are not needed to follow it. Figure~\ref{fig:examples} shows that the vorticities of single tilings look disordered [Figs.~\ref{fig:examples}(a) and \ref{fig:examples}(b)], whereas their average forms the $2\times2$ superstructure of Sec.~\ref{sec:order} [Figs.~\ref{fig:examples}(c) and \ref{fig:examples}(d)]; the total vorticity $V$ given in each panel is the same for all tilings of a sample (Sec.~\ref{sec:height}). Figure~\ref{fig:rect} tests the conformal-radius law, Eq.~\eqref{eq:amplaw}: for three aspect ratios the central amplitude follows a single fit in $H\rho$ [Fig.~\ref{fig:rect}(a)], and at fixed short side elongation lowers it by at most 4\% [Fig.~\ref{fig:rect}(b)]. Figures~\ref{fig:psi}(a)--\ref{fig:psi}(d) show the exact central phase field for the four residue classes of $n$; the phase is uniform and changes with $n\bmod4$ as Eq.~\eqref{eq:center} predicts (Sec.~\ref{sec:center}). In Figs.~\ref{fig:psi}(e)--\ref{fig:psi}(h) a monomer pair makes the phase wind by $\pi$ around each monomer, and opposite hues meet along the string of Sec.~\ref{sec:defects}. Figure~\ref{fig:holes} illustrates Sec.~\ref{sec:holes}: when the rim pattern of an $H=8$ hole agrees with the outer boundary ($k=0$), the amplitude rises toward the rim [Figs.~\ref{fig:holes}(a) and \ref{fig:holes}(d)]; when it disagrees ($k=+1$), the alignment $\cos\Delta\varphi$ changes sign across a crossover ring [Figs.~\ref{fig:holes}(b) and \ref{fig:holes}(e)]; for $H=10$ the two sectors are degenerate [Fig.~\ref{fig:holes}(f)].

\section{Kasteleyn methods}
\label{app:kasteleyn}

We use the Kasteleyn matrix $K$ with weights 1 (horizontal) and $i$ (vertical) between black and white cells~\cite{Kasteleyn1961,KenyonLectures}; for exact rational arithmetic, an equivalent real signing is combined with fraction-free (Bareiss) elimination. $\langle\eps(p)\rangle$ follows from Eq.~\eqref{eq:seam} with the seam flips placed on the edges crossed by a breadth-first lattice path. For a domino $d$ covering the black cell $b$ (row index $i$ of $K$) and the white cell $w$ (column index $j$), $\langle\mathbb 1_d\,\eps(p_d)\rangle=\eps_0K_\gamma(b,w)(-1)^{i+j}\det K_\gamma^{(bw)}/\det K$, where $\mathbb 1_d$ indicates that $d$ is present and $K_\gamma^{(bw)}$ is the minor without row $i$ and column $j$; this form also covers the cases in which $\det K_\gamma=0$.

\begin{figure*}[t]
\centering
\includegraphics[width=0.92\textwidth]{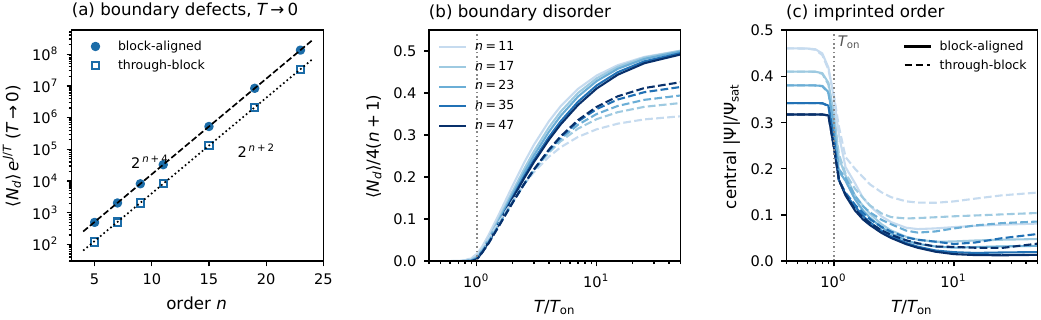}
\caption{Above $T_{\rm on}$ boundary defects disorder the edges of straight-edged arrays and erase the imprinted order (exact Pfaffian results, no statistical errors). Boundary defects are paired edge plaquettes and occupied half-dimer sites. (a) Low-temperature weight of the defects, $\langle N_d\rangle e^{J/T}$, versus the order $n$, for block-aligned (filled) and through-block (open) arrays; the lines are $2^{n+4}$ and $2^{n+2}$. (b) Defects per edge cell and (c) central amplitude $|\Psi|/\Psi_{\rm sat}$ (cells within distance 4 of the center, sign referred to the bottom corner) versus $T/T_{\rm on}$ for $n=11$--47; solid lines are block-aligned arrays with $T_{\rm on}=J/[(n+4)\ln2]$ [Eq.~\eqref{eq:Ton}], dashed lines through-block arrays with $T_{\rm on}=J/[(n+2)\ln2]$.}
\label{fig:edgeT}
\end{figure*}

\emph{Holes and defects.} For diagonal tiles the corresponding cells are removed and the shared corner of each diagonal tile is excluded from the paths. For a pair of monomers the single-cell holes violate the Kasteleyn condition; negating the edges crossed by a dual path between the two holes restores it, and a straight branch cut between the monomers is excluded from the seam paths. All cases were validated against brute-force enumeration (pairs of diagonal tiles at $n=4$--6, monomer pairs at $n=5$ and 6) to about $10^{-15}$. Pair weights $Z_{\rm pair}(r)/Z$ are ratios of determinants with and without the holes. The mean height follows from the edge probabilities $P(b,w)=K(b,w)K^{-1}(w,b)$ integrated along paths. The two-point function uses a seam from $p$ to $q$ [Eq.~\eqref{eq:twopoint}]. In double precision the relative error of $\langle\eps(\text{center})\rangle$, estimated against Eq.~\eqref{eq:product}, is below $10^{-4}$ for $n\le48$ and grows rapidly beyond; the largest sizes ($n\le58$) are used only on the logarithmic scales of Fig.~\ref{fig:exactscaling}. The exact rationals were obtained up to $n=31$.

\section{Simulations}
\label{app:sim}

Uniform Aztec-diamond tilings were generated by domino shuffling ($n\le195$, $10^3$--$2\times10^3$ tilings per size), and square and rectangular boards by height-function Metropolis flips (burn-in $20L^2$ sweeps, measurements spaced by $L^2/16$ sweeps, four independent replicas). Structure factors were computed by fast Fourier transform of $v$ placed on the plaquette centers. Dipolar simulations used Metropolis flips whose energy change was evaluated from the template spins of the two flipped tetrominoes with precomputed neighbor lists; the accumulated energy was checked against full recomputation. String lengths were obtained by integer linear programming of the minimal number of violated couplings, and vertex-model energies by integer programming over all island spins with vertex-type costs (unhappy $z=3$: 1; type-II and charged $z=4$, charged $z=2$ and $|q|=3$ $z=3$ vertices: $\Delta$; $\Delta=3$ unless stated).

\emph{Boundary excitations at finite temperature.} For the straight-edged samples of Sec.~\ref{sec:axis} the boundary monomers are the freeable edge plaquettes, with weight $e^{J/2T}$, and the half-dimer sites, with weight $n_a e^{-J/2T}$ (dimers have weight 1). Here $n_a$ counts the boundary vertices at which plaquette $a$ can carry a half dimer: 2 for a vertex of coordination 3 that touches only this plaquette, 1 for a corner vertex, and, for a vertex shared by two non-adjacent plaquettes, 1 for each of them that is not freeable. A plaquette can be of both kinds, and its weight $\ell_a$ is then the sum. All monomer sites lie on the outer face, so the partition function is $Z=\mathrm{Pf}(K+\Lambda)$~\cite{Giuliani2016}, with $K$ the real Kasteleyn matrix of the full box (horizontal edges $+1$, vertical edges $\pm1$ alternating by column), the sites numbered in cyclic order along the boundary, and $\Lambda_{ab}=-\Lambda_{ba}=G_aG_b\ell_a\ell_b$ for $a<b$, where $G_a$ is the product of Kasteleyn signs along the boundary from site 1 to site $a$. For the edge plaquettes alone we checked this against brute-force sums over all paired subsets for $n=5$ and 7 (agreement to $10^{-15}$) and against Monte Carlo sampling for $n=11$. The weights $n_a$ follow from a complete enumeration of the first excited level of the open vertex model; with them the exact density of states of the vertex model, obtained by a transfer matrix over all spin states, agrees with $Z$ at the first excited level for both terminations ($n=3$: 3584 and 768 states, $n=5$: 8\,126\,464 and 1\,966\,080 states above $2^{n(n+1)/2}$ ground states, block-aligned and through-block). At the second level the vertex model has 12\% (block-aligned) and 30\% (through-block) more states for $n=5$, down from 38\% and 150\% for $n=3$; these are bulk excitations, which cost at least $2J$. Monomer probabilities follow from $\Lambda_{ab}(K+\Lambda)^{-1}_{ba}$, and $\langle\eps(p)\rangle$ from the seam of Sec.~\ref{sec:exact}, evaluated as a low-rank Pfaffian update; the reference vertex is the bottom corner of the Aztec diamond. The results are shown in Fig.~\ref{fig:edgeT}.

\emph{Defect details.} On a $22\times22$ board (eight tilings per separation) the minimal number of violated couplings between two monomers grows by about 1.25 per chessboard step. On a $24\times24$ torus with the unhappy vertices fixed by the tiling, the excess energy of a monomer pair grows by one type-II $z=4$ vertex per diagonal step for $\Delta=2$, 3, and 5. Exact pair weights on a $64\times64$ board, obtained from Kasteleyn determinants with the defect cells removed, decay as $r^{-1/2}$ for monomers and as $r^{-2.05}$ for diagonal tiles counted together with their shared corner.

\emph{Dipolar details.} Moments were point dipoles or dumbbells of length $0.7a$, with a cutoff of $6a$ (results unchanged at $10a$). The cheapest excitation out of the ground-state manifold costs $3.8D$ (point) and $7.5D$ (dumbbell). Each size $L=24$, 32, 48, 64 was annealed in four independent replicas ($3\times10^3$ equilibration sweeps and 400 measurements spaced by 10 sweeps per temperature); errors are standard errors over replicas, and $2x$ is fitted to $\langle|\bar\Psi|^2\rangle\propto L^{-2x}$ in the central window with bootstrap errors. The Binder cumulant $U=1-\langle|c|^4\rangle/2\langle|c|^2\rangle^2$ of the crystal order parameter $c$, the $(\pi/2,\pi/2)$ Fourier component of the domino-orientation field, is close to 0 at high temperature and reaches $1/2$ below $T_c$; the curves for $L=32$--64 cross between $0.43D$ and $0.47D$. Above $T_c$, $U$ has a negative minimum, about $-0.4$ at $0.55D$ for $L=48$ and 64, whose depth does not decrease with $L$; this suggests a weakly first-order transition, which larger sizes would have to confirm. The specific heat peaks at $T_c$. The low-temperature crystal is neither columnar nor staggered and has several degenerate variants with different locked heights.

\clearpage
\bibliography{refs}

\end{document}